\documentclass[final,authoryear,5p,times,twocolumn]{elsarticle} % 用于生产投稿必须的Manuscript
\usepackage{multirow,setspace,times,amssymb,amsmath,graphicx,color,rotating,subfigure,url}
\usepackage{lineno}
\usepackage{dcolumn,multirow}
\usepackage{natbib}
\usepackage{pdflscape}

\hfuzz=\maxdimen
\begin{document}

\begin{frontmatter}

\title{Stylized facts of price gaps in limit order books: Evidence from Chinese stocks}
\author[SB]{Gao-Feng Gu}
\author[CME,CCSCA]{Xiong Xiong}
\author[CME,CCSCA]{Yong-Jie Zhang}
\author[SZSE]{Wei Chen}
\author[CME,CCSCA]{Wei Zhang}
\author[SB,SS]{Wei-Xing Zhou\corref{cor}}
\cortext[cor]{Corresponding author. Address: 130 Meilong Road, P.O. Box 114, School of Business, East China University of Science and Technology, Shanghai 200237, China, Phone: +86 21 64253634, Fax: +86 21 64253152.}
\ead{wxzhou@ecust.edu.cn} %

\address[SB]{School of Business, East China University of Science and Technology, Shanghai 200237, China}
%\address[RCE]{Research Center for Econophysics, East China University of Science and Technology, Shanghai 200237, China}
\address[CME]{College of Management and Economics, Tianjin University, Tianjin 300072, China}
\address[CCSCA]{China Center for Social Computing and Analytics, Tianjin University, Tianjin 300072, China}
\address[SZSE]{Shenzhen Stock Exchange, 5045 Shennan East Road, Shenzhen 518010, China}
\address[SS]{Department of Mathematics, East China University of Science and Technology, Shanghai 200237, China}

\begin{abstract}
  Price gap, defined as the logarithmic price difference between the first two occupied price levels on the same side of a limit order book (LOB), is a key determinant of market depth, which is one of the dimensions of liquidity. However, the properties of price gaps have not been thoroughly studied due to the less availability of ultrahigh frequency data. In the paper, we rebuild the LOB dynamics based on the order flow data of 26 A-share stocks traded on the Shenzhen Stock Exchange in 2003. Three key empirical statistical properties of price gaps are investigated. We find that the distribution of price gaps has a power-law tail for all stocks with an average tail exponent close to 3.2. Applying modern statistical methods, we confirm that the gap time series are long-range correlated and possess multifractal nature. These three features vary from stock to stock and are not universal. Furthermore, we also unveil buy-sell asymmetry phenomena in the properties of price gaps on the buy and sell sides of the LOBs for individual stocks. These findings deepen our understanding of the dynamics of liquidity of common stocks and can be used to calibrate agent-based computational financial models.
%
%  \medskip
  \noindent{\textit{JEL classification: G10, C14}}
\end{abstract}

\begin{keyword}
  Price gap, Limit order book, Liquidity, Stylized facts, Buy-sell asymmetry
%\PACS 89.65.Gh, 05.45.Tp, 89.75.Da
\end{keyword}

\end{frontmatter}

\section{Introduction}
\label{se:introduction}

It is a key stylized fact that returns over small time scales have power-law tails \citep{Mandelbrot-1963-JB,Mantegna-Stanley-1995-Nature,Gopikrishnan-Plerou-Amaral-Meyer-Stanley-1999-PRE,Plerou-Gopikrishnan-Amaral-Meyer-Stanley-1999-PRE,Bertram-2004-PA,CoronelBrizio-HernandezMontoya-2005-PA,Zhang-Zhang-Kleinert-2007-PA,Pan-Sinha-2008-PA,Tabak-Takami-Cajueiro-Petitiniga-2009-PA,Mu-Zhou-2010-PRE,Yang-Wang-Hu-2013-PLA,Liang-Yang-Huang-2013-FoP}, which means that large price changes occur more frequently than normal distribution. This phenomenon has important application in the domain of risk management, and it is necessary to understand the origins of large price fluctuations.

It is well-documented that trading volume is a key determinant to move stock prices. The relationship between price fluctuation and trading volume over certain time period has been extensively studied \citep{Karpoff-1987-JFQA}. There is numerous evidence showing that the magnitude of price fluctuation positively correlates to the trading volume at different time scales from one minute to one month \citep{Wood-McInish-Ord-1985-JF,Jain-Joh-1988-JFQA,Ying-1966-Em,Epps-1977-JFQA,Harris-1987-JFQA,Gallant-Rossi-Tauchen-1992-RFS,Richardson-Sefcik-Thompson-1986-JFE,Rogalski-1978-RES,Saatcioglu-Starks-1998-IJF}. The price-volume relation is usually asymmetric at the aggregate level in the sense that the price impact of a selling volume is larger than a buying volume of the same size \citep{Karpoff-1987-JFQA}. At the transaction level, theoretical and empirical analyses show that the price impact function is nonlinear \citep{Loeb-1983-FAJ,Perold-Salomon-1991-FAJ,Zhang-1999-PA,Farmer-2002-ICC,Almgren-2003-AMF,Gabaix-Gopikrishnan-Plerou-Stanley-2003-Nature,Gabaix-Gopikrishnan-Plerou-Stanley-2003-PA,Lillo-Farmer-Mantegna-2003-Nature,Farmer-Lillo-2004-QF,Plerou-Gopikrishnan-Gabaix-Stanley-2004-QF,Lim-Coggins-2005-QF,Gabaix-Gopikrishnan-Plerou-Stanley-2006-QJE,Gabaix-Gopikrishnan-Plerou-Stanley-2007-JEEA,Zhou-2012-NJP,Zhou-2012-QF}. At the transaction level, there is no buy-sell asymmetry in the price impact function \citep{Zhou-2012-QF}.

Trading volume or trade size is certainly not the solo driving force of price fluctuations. \cite{Farmer-Gillemot-Lillo-Mike-Sen-2004-QF} find that large price fluctuations of stocks traded on the London Stock Exchange are essentially independent of the volume of orders, but rather driven by liquidity fluctuations characterized by the gaps between the first few occupied price levels on the opposite limit order book. After investigating the TAQ data and order book data from the Island ECN, \cite{Weber-Rosenow-2006-QF} argue that a large trading volume alone is not sufficient to explain large price changes and a lack of liquidity is a necessary prerequisite for the occurrence of large price fluctuations. \cite{Naes-Skjeltorp-2006-JFinM} study the order flow data from the Oslo Stock Exchange and find that price fluctuations are positively correlated with trade number, a component of trading volume, and negatively correlated with different liquidity measures. \cite{Joulin-Lefevre-Grunberg-Bouchaud-2008-Wilmott} analyze the one-minute data of 163 USA stocks and find that news and trading volume play a minor role in causing large price changes. They conjecture that large price fluctuations are caused by the vanishing of liquidity. Based on the order flow data of Chinese stocks, \cite{Zhou-2012-NJP} finds that trade size, bid-ask spread, price gaps, and outstanding volumes all play a significant role in driving price fluctuations.

Bid-ask spread, price gap and standing volume on the LOBs are all fundamental ingredients of liquidity. The statistical properties of bid-ask spreads and volumes have been investigated for many financial markets \citep{Chakraborti-Toke-Patriarca-Abergel-2011a-QF,Gould-Porter-Williams-McDonald-Fenn-Howison-2013-QF}. However, only a few studies concern with the statistical properties of price gaps in financial markets. \cite{Farmer-Gillemot-Lillo-Mike-Sen-2004-QF} analyze the probability distribution of price gaps of a few stocks traded on London Stock Exchange and find that price gaps approximately follow a power-law distribution with the tail exponents varying from about 1.6 to 2.8. \cite{Lallouache-Abergel-2013-arXiv} focus on the EUR/USD and USD/JPY foreign exchange data from the Electronic Broking Service (EBS) Spot platform. They study the relation between the average gaps (in units of ticks) and price levels in the LOB and find that decimalized gaps decrease with the price levels in both buy and sell LOBs which do not change with time.

In this work, based on the order flow data of 26 A-share stocks traded on the Shenzhen Stock Exchange, we rebuild the LOBs according to the continuous double auction mechanism. We study the empirical statistical properties of the price gaps on the buy and sell LOBs. The rest of this paper is organized as follows. Section~\ref{se:dateset} briefly introduces the database we analyze. In Section~\ref{se:PDF}, we investigate the probability distributions of price gaps. Section~\ref{se:memory} estimates the memory effect of gap series using advanced statistical methods. We further investigate its multifractal nature in Section~\ref{se:multifractal}. Finally, we summarize the results in Section~\ref{se:conclusion}.

\section{Dateset}
\label{se:dateset}

Our study is based on the order flow data of 26 liquid stocks traded on the Shenzhen Stock Exchange, covering the whole year of 2003. The Shenzhen Stock Exchange adopts the continuous double auction mechanism, which was established on December 1, 1990 and started its operation on July 3, 1991. There are two kinds of independent markets on the SZSE, i.e., A-share market and B-share market. Both of them are open to mainland Chinese companies. The A-share market is traded in CNY and restricted to domestic investors, while the B-share market is traded in HKD and only open to foreign investors before February 19, 2001 since when it has been open to the domestic investors as well. Each A-share stock forms its open price through the call auction mechanism and enters the continuous double auction period since 9:30 in the morning. We focus on the data in the continuous double auction period.

There are 26 A-share stocks in our analysis, including Ping An Bank Co., Ltd. (000001), China Baoan Group Co., Ltd. (000009), CSG Holding Co., Ltd. (000012), Konka Group Co., Ltd. (000016), Shenzhen Kaifa Technology Co., Ltd. (000021), China Merchants Property Development Co., Ltd. (000024), Great Wall Computer Shenzhen Co., Ltd. (000066), Sinopec Shengli Oil Field Dynamic Group Co., Ltd. (000406), Guangdong Provincial Expressway Development Co., Ltd. (000429), Shandong Chenming Paper Holdings Co., Ltd. (000488), Guangdong Electric Power Development Co., Ltd. (000539), Foshan Electrical and Lighting Co., Ltd. (000541), Jiangling Motors Co., Ltd. (000550), Weifu High-Technology Group Co., Ltd. (000581), Chongqing Changan Automobile Co., Ltd. (000625), Hebei Iron and Steel Co., Ltd. (000709), Xinxing Ductile Iron Pipes Co., Ltd. (000778), Faw Car Co., Ltd. (000800), Shanxi Taigang Stainless Steel Co., Ltd. (000825), Citic Guoan Information Industry Co., Ltd. (000839), Wuliangye Yibin Co., Ltd. (000858), Angang Steel Co., Ltd. (000898), Hunan TV and Broadcast Intermediary Co., Ltd. (000917), Hunan Valin Steel Co., Ltd. (000932), Sinopec Zhongyuan Petroleum Co., Ltd. (000956), and Shanxi Xishan Coal and Electricity Power Co., Ltd. (000983).

The database records the order flows of the aforementioned stocks in 2003. It contains the details of order placement and order cancellation, including the order submitting time, order price, order size and order identifier which identifies whether the submitted order is a buy order, a sell order, or a cancelation. The time stamp is accurate to 0.01 second. We rebuilt the order book with respect to the placed orders and and cancelled orders according to the price-time priority rule \citep{Gu-Chen-Zhou-2008a-PA,Gu-Chen-Zhou-2008b-PA,Gu-Chen-Zhou-2008c-PA}. At each event time defined as a submitted or a cancelled order, we obtain the buy-side and sell-side LOBs, on which unexecuted limit orders occupy different price levels at $a_1$, $a_2$, $a_3$, $\cdots$ from low to high on the sell LOB and $b_1$, $b_2$, $b_3$, $\cdots$ from high to low on the buy LOB. The price gap $g(t)$ investigated in this work is defined as the absolute logarithmic difference between the first occupied price level (best bid or best ask) and the second occupied price level on the buy or sell LOB:
\begin{equation}
  g(t)=\left\{
  \begin{array}{ccc}
     \ln{b_1(t)}-\ln{b_2(t)}~~~{\mathrm{for~buy~LOB}}\\
     \ln{a_2(t)}-\ln{a_1(t)}~~~{\mathrm{for~sell~LOB}}
  \end{array}
  \right..
  \label{Eq:def:gap}
\end{equation}
Table \ref{Tb:Statistics} presents the basic statistics of the price gaps.

The second column and the ninth column of Table \ref{Tb:Statistics} show the order flow rate $\mu$ defined as the number of submitted order per minute. It is observed that the order flow rate varies remarkably from stock to stock, and not surprisingly $\mu_{b}$ strongly correlates with $\mu_{a}$. We further find that $\mu_{b}>\mu_{a}$ for 4 stocks and $\mu_{b}<\mu_{a}$ for 22 stocks. The third and tenth columns present the ratio $\omega$ of the number of gaps equaling to the tick size (0.01 CNY) to the total number of gaps for each stock. The value of $\omega$ varies from 0.73 to 0.99 for buy LOBs and from 0.70 to 0.99 for sell LOBs. It is not unexpected that $\omega_b$ correlates strongly with $\omega_a$. A closer scrutiny unveils that $\omega_b>\omega_a$ for 24 stocks.

The rest columns give the mean, median, standard deviation, skewness, and kurtosis of the gaps on each side of the LOB for each stock. These statistics vary from stock to stock. We observe strong correlations between the corresponding means of the price gaps on the buy and sell LOBs, so do the medians. Such buy-sell correlations are much weaker for other statistics. There are 22 stocks with larger mean gaps on the buy side. All the gap distributions are right skewed as expected because the mean gaps span less than two ticks. There are 24 stocks whose skewness of gaps on the buy LOB is greater than that on the sell LOB. We also find that all the gap distributions have large kurtosis and there are 23 stocks with greater kurtosis of the gaps on the buy LOBs.

\begin{landscape}
\begin{table}[htp]
  \centering
  \caption{Summary statistics of price gaps on the buy and sell LOBs of 26 stocks. $\mu$ is the order flow rate, i.e., the number of order submission per minute. $\omega$ is the ratio of the number with the gaps equaling to the tick size (0.01 CNY) to the total number of gaps on the buy LOB or sell LOB.}
  \medskip
  \label{Tb:Statistics}
  \centering
  \begin{tabular}{ccccccccccccccccccccccccccc}
  \hline\hline
       & \multicolumn{6}{c}{Buy LOB} && \multicolumn{6}{c}{Sell LOB} \\
  \cline{2-8} \cline{10-16}
  Stock & $\mu$ & $\omega$ & mean & median & s.t.d. & skewness & kurtosis & & $\mu$ & $\omega$ & mean & median & s.t.d. & skewness & kurtosis\\
  \hline
    000001 & 34.31 & 0.95 & 0.00098 & 0.00089 & 0.00036 &  8.77 &  164.4 && 30.95 & 0.94 & 0.00100 & 0.00089 & 0.00043 &  8.47 &  145.1 \\
    000009 & 17.74 & 0.98 & 0.00201 & 0.00194 & 0.00041 & 13.84 &  456.1 && 18.96 & 0.98 & 0.00201 & 0.00193 & 0.00045 &  9.53 &  232.1 \\
    000012 & 10.12 & 0.84 & 0.00123 & 0.00095 & 0.00088 &  9.00 &  178.2 &&  9.52 & 0.81 & 0.00129 & 0.00097 & 0.00090 &  4.89 &   44.2 \\
    000016 &  6.01 & 0.88 & 0.00151 & 0.00128 & 0.00087 & 15.94 &  672.5 &&  6.39 & 0.85 & 0.00157 & 0.00128 & 0.00098 & 11.37 &  357.2 \\
    000021 & 14.20 & 0.82 & 0.00099 & 0.00081 & 0.00062 &  5.13 &   54.6 && 14.73 & 0.79 & 0.00103 & 0.00081 & 0.00069 &  4.88 &   49.8 \\
    000024 &  4.09 & 0.80 & 0.00129 & 0.00098 & 0.00097 &  8.84 &  189.5 &&  4.87 & 0.79 & 0.00133 & 0.00098 & 0.00104 &  5.68 &   56.8 \\
    000066 &  9.78 & 0.86 & 0.00120 & 0.00100 & 0.00074 & 21.62 & 2110.5 && 10.35 & 0.84 & 0.00125 & 0.00101 & 0.00077 &  5.46 &   59.5 \\
    000406 &  9.67 & 0.94 & 0.00132 & 0.00122 & 0.00052 & 18.10 &  791.8 &&  9.54 & 0.93 & 0.00134 & 0.00122 & 0.00055 & 13.38 &  673.7 \\
    000429 &  3.86 & 0.94 & 0.00201 & 0.00187 & 0.00086 & 27.12 & 2217.2 &&  4.26 & 0.92 & 0.00206 & 0.00187 & 0.00095 &  7.08 &   93.9 \\
    000488 &  3.70 & 0.75 & 0.00122 & 0.00103 & 0.00107 & 12.02 &  280.9 &&  3.79 & 0.75 & 0.00125 & 0.00102 & 0.00095 &  4.36 &   34.6 \\
    000539 &  3.62 & 0.75 & 0.00143 & 0.00098 & 0.00144 & 11.33 &  295.4 &&  3.46 & 0.75 & 0.00142 & 0.00098 & 0.00122 &  4.97 &   45.3 \\
    000541 &  2.17 & 0.78 & 0.00123 & 0.00088 & 0.00122 & 14.18 &  371.5 &&  2.42 & 0.72 & 0.00134 & 0.00088 & 0.00127 & 12.40 &  411.1 \\
    000550 & 11.18 & 0.84 & 0.00116 & 0.00096 & 0.00065 &  5.37 &   65.5 && 12.08 & 0.81 & 0.00121 & 0.00097 & 0.00075 &  5.01 &   50.0 \\
    000581 &  3.03 & 0.75 & 0.00133 & 0.00092 & 0.00117 &  9.68 &  337.3 &&  3.68 & 0.71 & 0.00142 & 0.00092 & 0.00137 & 14.15 &  637.5 \\
    000625 & 11.89 & 0.75 & 0.00108 & 0.00087 & 0.00090 & 17.44 & 1022.5 && 12.51 & 0.72 & 0.00115 & 0.00088 & 0.00092 &  5.65 &   94.6 \\
    000709 &  7.28 & 0.97 & 0.00185 & 0.00191 & 0.00061 & 11.30 &  274.3 &&  7.66 & 0.97 & 0.00185 & 0.00191 & 0.00056 &  7.42 &  134.0 \\
    000778 &  4.88 & 0.86 & 0.00111 & 0.00090 & 0.00074 &  9.41 &  214.5 &&  5.86 & 0.84 & 0.00112 & 0.00090 & 0.00069 &  6.37 &   94.5 \\
    000800 & 17.81 & 0.90 & 0.00125 & 0.00109 & 0.00057 &  7.05 &  113.8 && 22.58 & 0.89 & 0.00126 & 0.00109 & 0.00058 &  5.33 &   57.9 \\
    000825 & 14.01 & 0.99 & 0.00205 & 0.00201 & 0.00033 & 10.84 &  226.7 && 15.98 & 0.98 & 0.00208 & 0.00200 & 0.00058 & 10.47 &  146.7 \\
    000839 & 20.08 & 0.83 & 0.00083 & 0.00065 & 0.00054 &  6.56 &  102.3 && 20.96 & 0.79 & 0.00087 & 0.00066 & 0.00064 &  6.06 &   77.2 \\
    000858 &  8.91 & 0.88 & 0.00112 & 0.00097 & 0.00058 &  7.64 &  140.3 &&  9.97 & 0.87 & 0.00114 & 0.00097 & 0.00063 &  6.72 &   94.8 \\
    000898 & 17.73 & 0.99 & 0.00232 & 0.00235 & 0.00042 &  6.55 &  170.3 && 24.14 & 0.99 & 0.00230 & 0.00234 & 0.00040 &  4.43 &   76.2 \\
    000917 &  5.98 & 0.73 & 0.00113 & 0.00080 & 0.00087 &  9.38 &  566.9 &&  6.25 & 0.70 & 0.00123 & 0.00081 & 0.00104 &  4.54 &   38.2 \\
    000932 & 11.49 & 0.99 & 0.00194 & 0.00189 & 0.00031 & 13.28 &  339.6 && 13.52 & 0.98 & 0.00194 & 0.00188 & 0.00040 & 26.16 & 1417.5 \\
    000956 & 12.72 & 0.91 & 0.00105 & 0.00094 & 0.00049 &  7.00 &   86.5 && 14.04 & 0.89 & 0.00106 & 0.00094 & 0.00051 &  6.56 &   86.4 \\
    000983 &  6.17 & 0.89 & 0.00144 & 0.00125 & 0.00074 & 13.33 &  981.2 &&  8.34 & 0.90 & 0.00143 & 0.00125 & 0.00069 &  5.53 &   61.0 \\
  \hline\hline
 \end{tabular}
\end{table}
\end{landscape}

\section{Probability distribution}
\label{se:PDF}

\begin{figure}[htb]
  \centering
  \includegraphics[width=8cm]{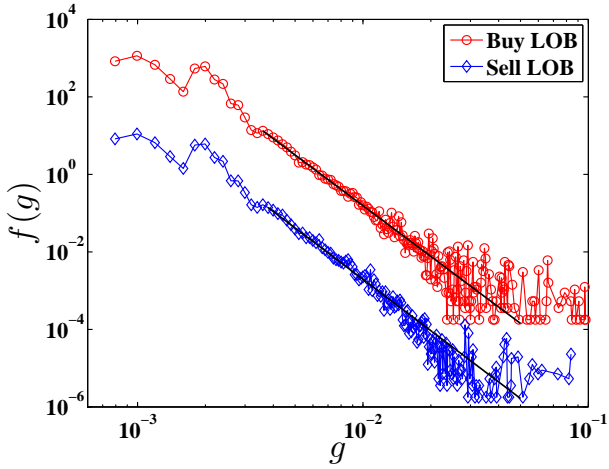}
  \caption{\label{Fig:PDF:26stocks} Empirical distributions $f(g)$ of ensemble gaps $g$ on the buy and sell LOBs for all the 26 stocks. The curve of sell LOB has been vertically translated for clarity. The solid lines are the power-law best fits to empirical data based on the KS test and MLE method.}
\end{figure}

Price gap is a significant determinant of immediate price impact \citep{Zhou-2012-NJP}, and the distribution of price gaps is directly related to the distribution of immediate price changes \citep{Farmer-Gillemot-Lillo-Mike-Sen-2004-QF}. In this section, we will investigate the distribution of gaps on both buy and sell LOBs for individual stocks. By looking at the empirical distributions of price gaps for the 26 stocks, we conjecture that these distributions have power-law tails:
\begin{equation}
 f(g) \sim g^{-\beta-1}~~~{\rm{for}}~~g \geq g_{\min},
 \label{Eq:PDF-f}
\end{equation}
where $\beta$ is the power-law exponent and $g_{\min}$ is the lower threshold of the scaling range of the power-law decay. We aggregate the gaps of all the stocks together and treat them as an ensemble. Figure~\ref{Fig:PDF:26stocks} illustrates the empirical probability distributions $f(g)$ of ensemble gaps on the buy LOB and the sell LOB. Evident power-law tails are observed. In addition, we see local humps in the distributions especially around $g=0.001$ and $g=0.002$. These humps are caused by the discreteness of the tick size and correspond to one and two ticks. These features are more evident for individual stocks.

To have a deeper understanding of the tail behavior, we need to conduct objective analysis. Based on the Kolmogorov-Smirnov test, \cite{Clauset-Shalizi-Newman-2009-SIAMR} propose an efficient quantitative method to test if the tail have a power-law form and, if so, to estimate the power-law exponent $\beta$ for the data greater than or equal to the threshold $g_{\min}$. We describe briefly the method of \cite{Clauset-Shalizi-Newman-2009-SIAMR}, which has been extensively applied in many fields. The Kolmogorov-Smirnov statistic ($KS$) is defined as
\begin{equation}
  KS=\max_{g \geq g_{\min}}(|P-F_{\rm{PL}}|),
  \label{Eq:KS}
\end{equation}
where $P$ is the cumulative distribution of gaps and $F_{\rm{PL}}$ is the cumulative distribution of the best power-law fit. The threshold $g_{\min}$ is determined by minimizing the $KS$ statistic. Then the power-law exponent $\beta$ of the data in the range $g \geq g_{\min}$ can be estimated using the maximum likelihood estimation (MLE) method, that is,
\begin{equation}
 \beta = n\left[\sum_{i=1}^{n}{\ln\frac{g(i)}{g_{\min}}}\right]^{-1},
 \label{Eq:beta}
\end{equation}
where $n$ is the number of the data points in the range $g>g_{\min}$. The standard error $\sigma$ on the power-law exponent $\beta$ is derived from the width of the likelihood maximum, which reads
\begin{equation}
 \sigma=\frac{\beta}{\sqrt{n}}.
 \label{Eq:err}
\end{equation}

Applying this approach to the ensemble data shown in Fig.~\ref{Fig:PDF:26stocks}, we have $\beta_{\rm{b}}=3.34$ with $g_{\min}=0.0035$ and $\sigma=0.0049$ for the buy LOB of ensemble gaps and $\beta_{\rm{s}}=3.39$ with $g_{\min}=0.0036$ and $\sigma=0.0046$. We also analyze individual stocks and present the minimal gaps, the tail exponents, and the standard deviations for each stock in Table~\ref{Tb:PDF-PL}. For buy LOBs, the value of $\beta_{\rm{b}}$ varies in the range $\left[2.38,4.35\right]$ with the mean value $\overline{\beta}_{\rm{b}}=3.19\pm0.53$. For sell LOBs, the tail exponent $\beta_{\rm{s}}$ fluctuates in the range $\left[2.16,4.60\right]$ with the mean value $\overline{\beta}_{\rm{s}}=3.17\pm0.60$. It is clear that the mean values $\overline{\beta}_{\rm{b}}$ and $\overline{\beta}_{\rm{s}}$ are similar to the power-law tail exponents obtained from ensemble gaps. However, the tail exponents of individual stocks have larger standard deviations. For stocks traded on the London Stock Exchange, \cite{Farmer-Gillemot-Lillo-Mike-Sen-2004-QF} find that the tail exponents vary from about 1.6 to 2.8 and the tail exponents of gaps and returns are almost identical. They thus argue that large price changes are mainly caused by price gaps. Although the tail exponents of gaps of Chinese stocks are systemically greater than those of British stocks, the tail exponents of gaps and returns of Chinese stocks are also very close \citep{Zhou-2012-QF}, seemingly supporting the conjecture of \cite{Farmer-Gillemot-Lillo-Mike-Sen-2004-QF}.

\begin{table*}[htp]
  \caption{Characteristic parameters in the power-law distributions of price gaps $g$ on both buy and sell cancellations of 26 stocks based on Kolmogorov-Smirnov tests and maximum likelihood estimation.}
  \medskip
  \label{Tb:PDF-PL}
  \centering
  \begin{tabular}{cccccccccccccc}
 \hline\hline
  & \multicolumn{5}{c}{Buy LOB} && \multicolumn{5}{c}{Sell LOB} \\
  \cline{2-6} \cline{8-12}
  Stock & $g_{\rm{min}}$ & $\beta_{\rm{b}}$ & $\sigma$ & $KS$ & $p$-value && $g_{\rm{min}}$ & $\beta_{\rm{b}}$ & $\sigma$ & $KS$ & $p$-value \\
  \hline
    000001 & 0.0025 & 3.16 & 0.018 & 0.057 & 1.00 && 0.0016 & 2.98 & 0.007 & 0.057 & 1.00  \\
    000009 & 0.0034 & 4.29 & 0.023 & 0.072 & 1.00 && 0.0035 & 4.52 & 0.020 & 0.082 & 1.00  \\
    000012 & 0.0026 & 2.64 & 0.012 & 0.049 & 1.00 && 0.0010 & 2.16 & 0.003 & 0.059 & 1.00  \\
    000016 & 0.0026 & 3.04 & 0.014 & 0.038 & 1.00 && 0.0022 & 3.11 & 0.010 & 0.041 & 1.00  \\
    000021 & 0.0016 & 3.05 & 0.007 & 0.051 & 1.00 && 0.0016 & 2.84 & 0.006 & 0.058 & 1.00  \\
    000024 & 0.0025 & 2.74 & 0.015 & 0.069 & 1.00 && 0.0026 & 2.68 & 0.013 & 0.074 & 1.00  \\
    000066 & 0.0029 & 3.29 & 0.018 & 0.025 & 1.00 && 0.0029 & 3.22 & 0.015 & 0.041 & 1.00  \\
    000406 & 0.0022 & 4.02 & 0.016 & 0.048 & 1.00 && 0.0022 & 4.03 & 0.014 & 0.062 & 1.00  \\
    000429 & 0.0032 & 3.55 & 0.021 & 0.059 & 1.00 && 0.0032 & 3.11 & 0.016 & 0.060 & 1.00  \\
    000488 & 0.0026 & 2.51 & 0.017 & 0.048 & 1.00 && 0.0028 & 2.97 & 0.018 & 0.048 & 1.00  \\
    000539 & 0.0026 & 2.38 & 0.012 & 0.053 & 1.00 && 0.0026 & 2.47 & 0.012 & 0.076 & 1.00  \\
    000541 & 0.0032 & 2.65 & 0.026 & 0.055 & 1.00 && 0.0032 & 3.06 & 0.024 & 0.046 & 1.00  \\
    000550 & 0.0022 & 3.04 & 0.011 & 0.051 & 1.00 && 0.0017 & 2.66 & 0.006 & 0.060 & 1.00  \\
    000581 & 0.0024 & 2.55 & 0.013 & 0.058 & 1.00 && 0.0024 & 2.49 & 0.011 & 0.060 & 1.00  \\
    000625 & 0.0023 & 2.86 & 0.010 & 0.021 & 1.00 && 0.0025 & 2.95 & 0.010 & 0.039 & 1.00  \\
    000709 & 0.0027 & 2.97 & 0.019 & 0.074 & 0.80 && 0.0027 & 3.26 & 0.019 & 0.079 & 0.99  \\
    000778 & 0.0025 & 2.92 & 0.018 & 0.045 & 1.00 && 0.0024 & 3.23 & 0.019 & 0.055 & 1.00  \\
    000800 & 0.0019 & 3.26 & 0.008 & 0.051 & 1.00 && 0.0028 & 3.64 & 0.015 & 0.040 & 1.00  \\
    000825 & 0.0043 & 3.41 & 0.048 & 0.131 & 0.00 && 0.0035 & 2.93 & 0.039 & 0.106 & 0.00  \\
    000839 & 0.0027 & 3.50 & 0.018 & 0.025 & 1.00 && 0.0027 & 3.42 & 0.015 & 0.046 & 1.00  \\
    000858 & 0.0027 & 3.54 & 0.022 & 0.053 & 1.00 && 0.0028 & 3.42 & 0.019 & 0.063 & 1.00  \\
    000898 & 0.0049 & 4.35 & 0.056 & 0.172 & 0.00 && 0.0049 & 4.60 & 0.069 & 0.175 & 0.00  \\
    000917 & 0.0014 & 2.50 & 0.006 & 0.044 & 1.00 && 0.0013 & 2.19 & 0.005 & 0.044 & 1.00  \\
    000932 & 0.0041 & 3.59 & 0.053 & 0.215 & 0.00 && 0.0042 & 3.26 & 0.046 & 0.226 & 0.00  \\
    000956 & 0.0026 & 3.52 & 0.019 & 0.055 & 1.00 && 0.0025 & 3.67 & 0.019 & 0.061 & 1.00  \\
    000983 & 0.0022 & 3.46 & 0.012 & 0.065 & 1.00 && 0.0022 & 3.40 & 0.012 & 0.057 & 1.00  \\
  \hline\hline
 \end{tabular}
\end{table*}

Following \cite{Clauset-Shalizi-Newman-2009-SIAMR}, we perform bootstrap test if the gap distributions have power-law tails. In doing so, we generate 100 realizations of gaps for each distribution. For each realization, we compute the KS statistic $KS_{\rm{sim}}$ as follows:
\begin{equation}
  KS_{\rm{sim}} = \max (|P_{\rm{sim}}-F_{\rm{PL}}|),
  \label{Eq:KS:sim}
\end{equation}
where $P_{\rm{sim}}$ is the cumulative distribution of the simulated realization. We thus obtain the $p$-value:
\begin{equation}
  p{\rm{-value}} = \frac{\#(KS_{\rm{sim}}>KS)}{100},
\end{equation}
where the numerator is the number of realizations with $KS_{\rm{sim}}>KS$. The meaning of this rest is that the investigated gaps have the power-law distribution with a probability of $p$. The resulting $KS$ values and the corresponding $p$-values are given in Table \ref{Tb:PDF-PL}. It is found that three stocks have a $p$-value of zero, while other stocks have very large $p$-values. This test confirms that most stocks have power-law tails in the gap distributions.

\begin{figure}[htb]
  \centering
  \includegraphics[width=8cm]{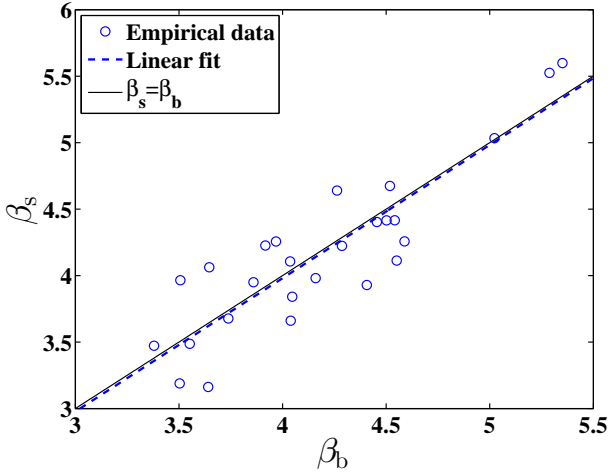}
  \caption{\label{Fig:PDF:betas} Linear relation between the power-law exponents of buy LOBs ($\beta_{{\rm{b}}}$) and sell LOBs ($\beta_{{\rm{s}}}$). The dash line is the fit to the data points with the robust regression and the solid line stands for the relation $\beta_{{\rm{s}}}=\beta_{{\rm{b}}}$.}
\end{figure}

We further illustrate the relation between $\beta_{{\rm{b}}}$ and $\beta_{{\rm{s}}}$ in Fig.~\ref{Fig:PDF:betas}, and quantitatively analyze the linear relation between them using the robust regression, which reads
\begin{equation}
 \begin{split}
  \beta_{{\rm{s}}}=&-0.0306~~+~~1.0027\beta_{{\rm{b}}}~, \\
  &~~(0.9497)~~~~~~~(0.0000) \\
 \end{split}
 \label{Eq:RL-beta}
\end{equation}
where the numbers in parentheses are the $p$-values of the coefficients. We find that the intercept is insignificantly different from zero, while the coefficient of $\beta_{{\rm{b}}}$ is evidently significant with the $p$-value close to zero. It is clear that the fitted dash line from the robust regressing method almost overlaps with the solid line $\beta_{{\rm{s}}}=\beta_{{\rm{b}}}$. Therefore, although there is certain buy-sell asymmetry in the tail distributions of gaps for individual stocks, the tail behaviors of gaps on the two LOBs for a same stock share great similarity.

\section{Long-range correlation}
\label{se:memory}

In this section, we investigate if there are long-range corrections in gap time series. There are many methods proposed to estimate the memory effect of time series \citep{Taqqu-Teverovsky-Willinger-1995-Fractals,Bashan-Bartsch-Kantelhardt-Havlin-2008-PA,Barunik-Kristoufek-2010-PA}, such as rescaled range (RS) analysis \citep{Hurst-1951-TASCE}, fluctuation analysis (FA) \citep{Peng-Buldyrev-Goldberger-Havlin-Sciortino-Simons-Stanley-1992-Nature}, wavelet transform module maxima (WTMM) \citep{Holschneider-1988-JSP,Muzy-Bacry-Arneodo-1991-PRL}, detrended fluctuation analysis (DFA) \citep{Peng-Buldyrev-Havlin-Simons-Stanley-Goldberger-1994-PRE}, and detrending moving average (DMA) \citep{Alessio-Carbone-Castelli-Frappietro-2002-EPJB}, to list a few. We adopt the DFA and DMA algorithms, which are among the most effective and the most extensively used methods \citep{Shao-Gu-Jiang-Zhou-Sornette-2012-SR}.

For a given price gap time series $\{g(t)|t =1,2,\cdots,N\}$, we calculate the cumulative summation series $G(t)$ as follows,
\begin{equation}
  G(t) = \sum_{j=1}^{t} \left[g(j)-\langle{g}\rangle\right],~~t = 1, 2, \cdots, N,
  \label{Eq:cumsum}
\end{equation}
where $\langle{g}\rangle$ is the sample mean of the $g(t)$ series. The series $G$ is covered by $N_s$ disjoint boxes with the same size $s$. When the whole series $G(t)$ cannot be completely covered by $N_s$ boxes, we can utilize $2N_s$ boxes to cover the series from both ends of the series. In each box, a trend function ${\tilde{G}}(t)$ of the sub-series is determined. The residuals are calculated by
\begin{equation}
  \epsilon(t) = G(t)-{\tilde{G}}(t).
  \label{Eq:epsilon}
\end{equation}
There are many different methods for the determination of $\tilde{G}$. The local detrending functions could be polynomials, which recovers the DFA method \citep{Peng-Buldyrev-Havlin-Simons-Stanley-Goldberger-1994-PRE,Hu-Ivanov-Chen-Carpena-Stanley-2001-PRE}. The local detrending function could also be the moving averages, resulting in the DMA algorithm \citep{Vandewalle-Ausloos-1998-PRE,Alessio-Carbone-Castelli-Frappietro-2002-EPJB,Xu-Ivanov-Hu-Chen-Carbone-Stanley-2005-PRE,Arianos-Carbone-2007-PA}.

The local fluctuation function $f_v(s)$ in the $v$-th box is defined as the r.m.s. of the residuals:
\begin{equation}
  \left[f_v(s)\right]^2 = \frac{1}{s}\sum_{t=(v-1)s+1}^{vs} \left[\epsilon(t)\right]^2~.
  \label{Eq:fv:s}
\end{equation}
The overall fluctuation function is calculated as follows:
\begin{equation}
  F(s) = \left\{\frac{1}{N_s}\sum_{v=1}^{N_s} {F_v^2(s)}\right\}^{\frac{1}{2}},
  \label{Eq:F2s}
\end{equation}
For most time series with fractal nature, one has:
\begin{equation}
 F(s) \sim s^{H},
 \label{Eq:H}
\end{equation}
where $H$ can be roughly viewed as the Hurst exponent. If $H$ is significantly greater than 0.5, the time series is positively correlated. If $H$ is insignificantly different from 0.5, the time series is uncorrelated. If $H$ is significantly smaller than 0.5, the time series is negatively correlated. When $H$ is compared with 0.5, statistical tests are necessary \citep{Jiang-Xie-Zhou-2014-PA}.

\begin{figure}[htb]
  \centering
  \includegraphics[width=8cm]{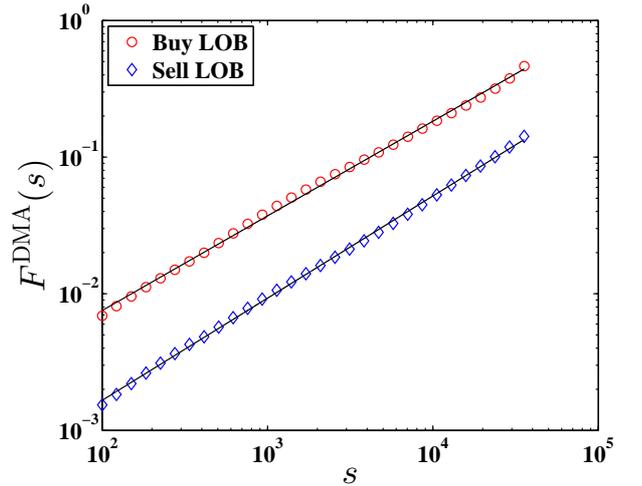}
  \caption{\label{Fig:DMA-FS} Detrending moving average analysis of the two gap time series for stock 000016. The solid lines are power-law fits to the data. The curve for the sell LOB has been shifted vertically for clarity.}
\end{figure}

We first choose the centred detrending moving average (CDMA) method, which has a better performance than the backward detrending moving average (BDMA) method and the forward detrending moving average (FDMA) method for the positively correlated time series \citep{Arianos-Carbone-2007-PA}. Figure~\ref{Fig:DMA-FS} presents the DMA detrended fluctuation functions $F_{\rm{DMA}}(s)$ with respect to the scale sizes $s$ for the gap series on both buy and sell LOBs of the stock 000016. It is obvious that the empirical data points all collapse to the fitting line, indicating an evident power-law scaling relation between the fluctuation function $F_{\rm{DMA}}(s)$ and the scale size $s$:
\begin{equation}
 F_{\rm{DMA}}(s) \sim s^{H^{\rm{DMA}}},
 \label{Eq:DMA-FS}
\end{equation}
where $H^{\rm{DMA}}$ is known as the DMA scaling exponent. Using the least squares fitting method, we obtain the scaling exponent $H_{\rm{b}}^{\rm{DMA}}=0.710 \pm 0.007$ for the buy LOB and $H_{\rm{s}}^{\rm{DMA}}=0.762 \pm 0.005$ for the sell LOB of the stock 000016.

We then calculate the DMA scaling exponents for the rest stocks, and the results are listed in Table~\ref{Tb:DMADFA-H}. The exponent $H_{\rm{b}}^{\rm{DMA}}$ varies in the range $[0.710,~0.797]$ with the mean value $\overline{H}_{\rm{b}}^{\rm{DMA}}=0.747\pm0.024$ for buy LOBs and the exponent $H_{\rm{s}}^{\rm{DMA}}$ fluctuates in the range $[0.719,~0.876]$ with the mean value $\overline{H}_{\rm{s}}^{\rm{DMA}}=0.768\pm0.029$ for sell LOBs. Since the DMA scaling exponents $H^{\rm{DMA}}$ are obviously larger than 0.5, we conclude that the gap series process long memory for both buy and sell LOBs of 26 stocks.

\begin{table*}[htp]
  \centering
  \caption{The scaling exponents qualifying long-range correlations in gap time series on the buy and sell LOBs of 26 stocks based on the DMA and DFA methods. $H_{\rm{SFL}}$ is the average scaling exponent of 100 shuffled gap time series.}
  \medskip
  \label{Tb:DMADFA-H}
  \centering
  \begin{tabular}{ccccccccccccc}
  \hline\hline
  && \multicolumn{5}{c}{Buy LOB} && \multicolumn{5}{c}{Sell LOB} \\
  \cline{3-7} \cline{9-13}
  Stock && $H_{\rm{b}}^{\rm{DMA}}$ & $H_{\rm{SFL}}^{\rm{DMA}}$ && $H_{\rm{b}}^{\rm{DFA}}$ & $H_{\rm{SFL}}^{\rm{DFA}}$ && $H_{\rm{s}}^{\rm{DMA}}$ & $H_{\rm{SFL}}^{\rm{DMA}}$ && $H_{\rm{s}}^{\rm{DFA}}$ & $H_{\rm{SFL}}^{\rm{DFA}}$ \\
  \hline
    000001 && 0.797 & 0.499 && 0.797 & 0.501 && 0.811 & 0.501 && 0.804 & 0.501 \\
    000009 && 0.773 & 0.500 && 0.750 & 0.500 && 0.786 & 0.500 && 0.763 & 0.499 \\
    000012 && 0.744 & 0.500 && 0.750 & 0.500 && 0.771 & 0.499 && 0.780 & 0.500 \\
    000016 && 0.710 & 0.502 && 0.714 & 0.500 && 0.762 & 0.499 && 0.765 & 0.501 \\
    000021 && 0.756 & 0.500 && 0.742 & 0.500 && 0.762 & 0.500 && 0.769 & 0.500 \\
    000024 && 0.717 & 0.503 && 0.737 & 0.500 && 0.756 & 0.499 && 0.762 & 0.501 \\
    000066 && 0.736 & 0.501 && 0.744 & 0.501 && 0.744 & 0.501 && 0.749 & 0.500 \\
    000406 && 0.732 & 0.500 && 0.739 & 0.500 && 0.768 & 0.501 && 0.774 & 0.500 \\
    000429 && 0.715 & 0.502 && 0.724 & 0.501 && 0.719 & 0.500 && 0.729 & 0.499 \\
    000488 && 0.757 & 0.499 && 0.772 & 0.500 && 0.773 & 0.499 && 0.788 & 0.500 \\
    000539 && 0.755 & 0.497 && 0.762 & 0.499 && 0.788 & 0.501 && 0.795 & 0.500 \\
    000541 && 0.740 & 0.499 && 0.739 & 0.499 && 0.751 & 0.498 && 0.761 & 0.500 \\
    000550 && 0.716 & 0.498 && 0.723 & 0.500 && 0.738 & 0.498 && 0.761 & 0.500 \\
    000581 && 0.737 & 0.499 && 0.745 & 0.500 && 0.773 & 0.500 && 0.785 & 0.499 \\
    000625 && 0.727 & 0.500 && 0.725 & 0.500 && 0.779 & 0.502 && 0.780 & 0.500 \\
    000709 && 0.754 & 0.501 && 0.732 & 0.500 && 0.762 & 0.500 && 0.743 & 0.499 \\
    000778 && 0.748 & 0.501 && 0.752 & 0.501 && 0.747 & 0.500 && 0.758 & 0.500 \\
    000800 && 0.747 & 0.500 && 0.754 & 0.500 && 0.773 & 0.500 && 0.781 & 0.500 \\
    000825 && 0.784 & 0.501 && 0.787 & 0.500 && 0.876 & 0.501 && 0.881 & 0.501 \\
    000839 && 0.766 & 0.501 && 0.767 & 0.501 && 0.775 & 0.500 && 0.777 & 0.501 \\
    000858 && 0.723 & 0.500 && 0.744 & 0.499 && 0.773 & 0.501 && 0.773 & 0.501 \\
    000898 && 0.791 & 0.501 && 0.727 & 0.501 && 0.749 & 0.500 && 0.697 & 0.499 \\
    000917 && 0.728 & 0.501 && 0.733 & 0.500 && 0.772 & 0.500 && 0.779 & 0.501 \\
    000932 && 0.752 & 0.500 && 0.750 & 0.500 && 0.744 & 0.501 && 0.742 & 0.498 \\
    000956 && 0.765 & 0.500 && 0.768 & 0.500 && 0.766 & 0.501 && 0.773 & 0.501 \\
    000983 && 0.755 & 0.499 && 0.756 & 0.500 && 0.735 & 0.502 && 0.730 & 0.501 \\
  \hline\hline
 \end{tabular}
\end{table*}

\begin{figure}[htb]
  \centering
  \includegraphics[width=8cm]{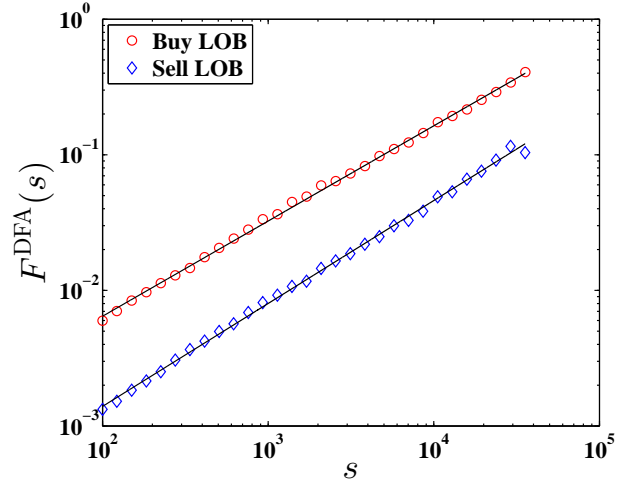}
  \caption{\label{Fig:DFA-FS} Detrended fluctuation analysis of the two gap time series for stock 000016. The solid lines are power-law fits to the data. The curve for the sell LOB has been shifted vertically for clarity.}
\end{figure}

We alternatively apply the detrended fluctuation analysis (DFA) method to confirm the long memory effect in gap series. Figure~\ref{Fig:DFA-FS} depicts the DFA detrended fluctuation functions $F_{\rm{DFA}}(s)$ as a function of the scale size $s$ for both buy and sell LOBs of the same stock 000016. We observe perfect power-law scaling relation between the fluctuation function $F_{\rm{DFA}}(s)$ and the scale size $s$:
\begin{equation}
 F_{\rm{DFA}}(s) \sim s^{H^{\rm{DFA}}}~,
 \label{Eq:DFA-FS}
\end{equation}
where $H^{\rm{DFA}}$ is DFA scaling exponent. Using the least-squares fitting method, we obtain $H_{\rm{b}}^{\rm{DFA}}=0.714 \pm 0.006$ for the buy LOB and $H_{\rm{s}}^{\rm{DFA}}=0.765 \pm 0.008$ for the sell LOB of stock 000016.

We also estimate the memory effect of the rest stocks with the DFA method. The DFA scaling exponents of 26 stocks are presented in Table~\ref{Tb:DMADFA-H} as well. We find that the scaling exponent $H_{\rm{b}}^{\rm{DFA}}$ varies from 0.714 to 0.797 with the mean value $\overline{H}_{\rm{b}}^{\rm{DFA}}=0.747\pm0.019$ for buy LOBs, and the exponent $H_{\rm{s}}^{\rm{DFA}}$ fluctuates from 0.697 to 0.881 with the mean value $\overline{H}_{\rm{s}}^{\rm{DFA}}=0.769\pm0.032$ for sell LOBs. Since the scaling exponents calculated from the two methods obviously larger than 0.5, It is believed that the gap series on both buy and sell LOBs of 26 stocks process long memory.

On the other hand, the probability distribution of gaps may affect its memory effect. In order to test the distribution effect, we first shuffle the gap series of each stock for 100 times, then obtain the shuffled scaling exponents $H_{\rm{SFL}}^{\rm{DMA}}$ and $H_{\rm{SFL}}^{\rm{DFA}}$ based on the DMA and DFA methods, respectively. The average values of 100 shuffled gap series are also illustrated in Table~\ref{Tb:DMADFA-H}. It is clear that the values of $H_{\rm{SFL}}^{\rm{DMA}}$ and $H_{\rm{SFL}}^{\rm{DFA}}$ for both buy and sell LOBs extremely approach to 0.5 which are obviously smaller than the original ones. So we make a conclusion that the probability distribution of gap series does not affect the memory effect, and confirm that gap series truly process significant long memory for all the 26 stocks.

\begin{figure}[htb]
  \centering
  \includegraphics[width=8cm]{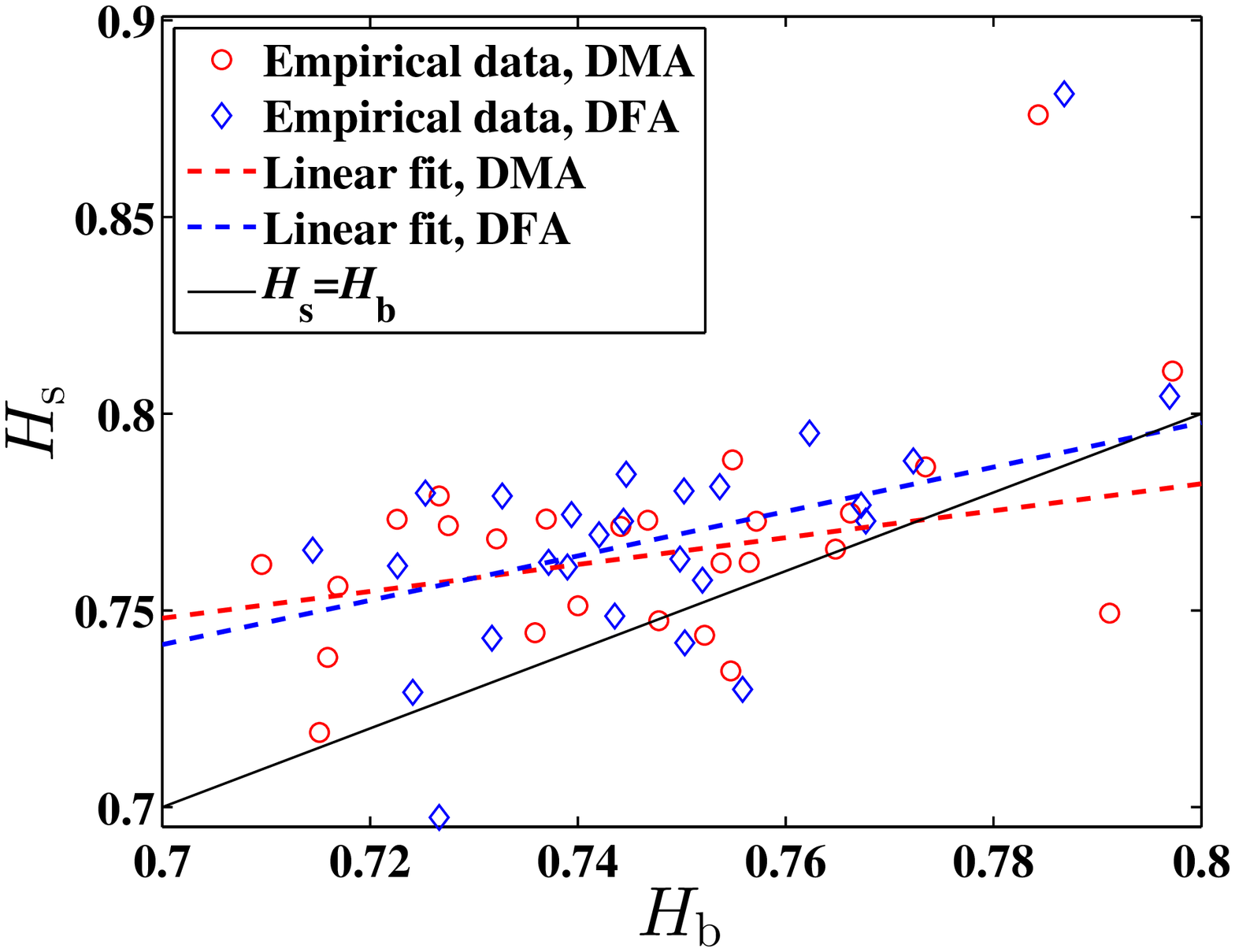}
  \includegraphics[width=8cm]{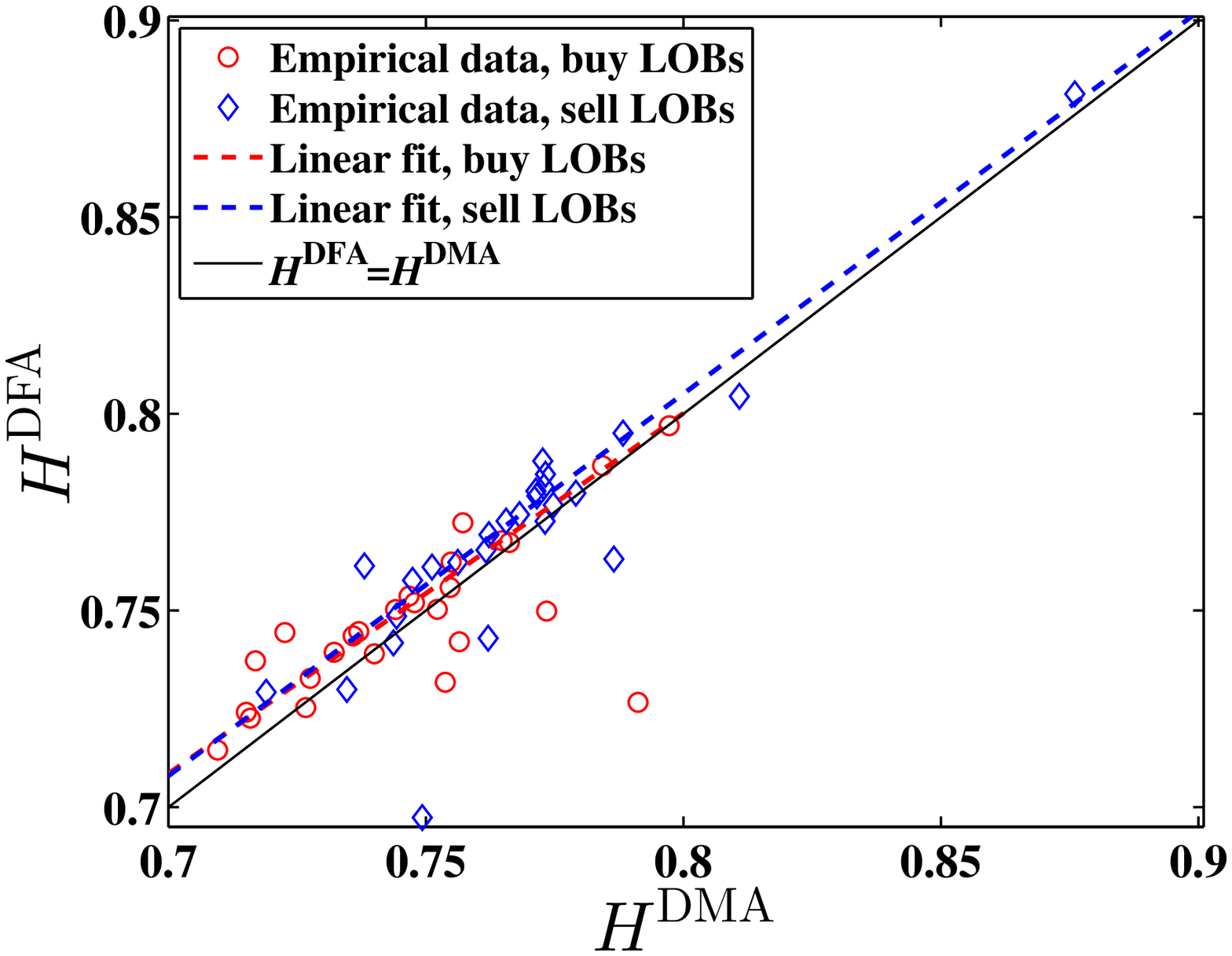}
  \caption{\label{Fig:RL-DMA-DFA} Panel (a): Plots of the relations between the scaling exponent of buy LOBs $H_{\rm{b}}$ and the ones of sell LOBs $H_{\rm{s}}$ with the DMA method and the DFA method. Panel (b): Plot of the relations between the DMA scaling exponent $H^{\rm{DMA}}$ and the DFA scaling exponent $H^{\rm{DFA}}$ for both buy and sell LOBs. The dash lines are linear fits to the empirical data, and the solid lines present the relations $H_{\rm{s}}=H_{\rm{b}}$ (a) and $H^{\rm{DFA}}=H^{\rm{DMA}}$ (b), respectively.}
\end{figure}

We then analyze the linear relation between $H_{\rm{b}}$ and $H_{\rm{s}}$ for all 26 stocks with DMA and DFA methods, which are showed in Fig.~\ref{Fig:RL-DMA-DFA}(a). With the robust regressing method, for the DMA method we have the following linear relation,
\begin{equation}
 \begin{split}
  H^{{\rm{DMA}}}_{{\rm{s}}}=&~~0.5087~~+~~0.3418H^{{\rm{DMA}}}_{{\rm{b}}}~, \\
  &~(0.0008)~~~~~(0.0659) \\
 \end{split}
 \label{Eq:Hurst-DMA}
\end{equation}
and for the DFA method we have
\begin{equation}
 \begin{split}
  H^{{\rm{DFA}}}_{{\rm{s}}}=&~~0.3464~~+~~0.5642H^{{\rm{DFA}}}_{{\rm{b}}}~, \\
  &~(0.0392)~~~~~(0.0138) \\
 \end{split}
 \label{Eq:Hurst-DFA}
\end{equation}
where the numbers in parentheses are the $p$-values of the coefficients. We find that the coefficients are significant at the 10\% level for the DMA method and at the 5\% level for the DFA method. According to Table~\ref{Tb:DMADFA-H} and Fig.~\ref{Fig:RL-DMA-DFA}(a), most stocks have $H_{\rm{s}}>H_{\rm{b}}$.

In order to compare the results obtained from the DMA method with the DFA method, we investigate the relation between the DMA scaling exponents $H^{{\rm{DMA}}}$ and the DFA scaling exponents $H^{{\rm{DFA}}}$ for both buy and sell LOBs of the 26 stocks. The results are presented in Fig.~\ref{Fig:RL-DMA-DFA}(b). We observe that $H^{{\rm{DMA}}}$ and $H^{{\rm{DFA}}}$ are linearly correlated. With the robust regression method, for buy LOBs we have
\begin{equation}
 \begin{split}
  H^{{\rm{DFA}}}_{{\rm{b}}}=&~~0.0677~~+~~0.9155H^{{\rm{DMA}}}_{{\rm{b}}}~, \\
  &~(0.1520)~~~~~(0.0000) \\
 \end{split}
 \label{Eq:Hurst-DMA2}
\end{equation}
and for sell LOBs we have
\begin{equation}
 \begin{split}
  H^{{\rm{DFA}}}_{{\rm{s}}}=&~~0.0277~~+~~0.9717H^{{\rm{DMA}}}_{{\rm{s}}}~, \\
  &~(0.5323)~~~~~(0.0000) \\
 \end{split}
 \label{Eq:Hurst-DFA2}
\end{equation}
where the numbers in parentheses are the $p$-values of the coefficients. In the regression equation the intercept is not significant for the two methods, while the coefficients of $H^{{\rm{DMA}}}$ and $H^{{\rm{DMA}}}$ are significant with the $p$-values close to zero. We conclude that both DMA and DFA methods unveil significant long-term correlations in the gap time series and both methods give quantitatively similar results.

\section{Multifractal nature}
\label{se:multifractal}

We now turn to study the possible presence of nonlinear correlations in the gap time series through multifractal analysis. We adopt the multifractal detrended fluctuation analysis proposed \cite{Kantelhardt-Zschiegner-KoscielnyBunde-Havlin-Bunde-Stanley-2002-PA}. We note that there are other methods that can be used for multifractal analysis \citep{DiMatteo-Aste-Dacorogna-2005-JBF}. We calculate the $q$th-order detrended fluctuation function as follow:
\begin{equation}
  F_q(s) = \left\{\frac{1}{N_s}\sum_{v=1}^{N_s} {F_v^q(s)}\right\}^{\frac{1}{q}},
  \label{Eq:Fqs}
\end{equation}
where $q$ can take any real value except for $q=0$. When $q=0$,
we have
\begin{equation}
  \ln[F_0(s)] = \frac{1}{N_s}\sum_{v=1}^{N_s}{\ln[F_v(s)]},
  \label{Eq:Fq0}
\end{equation}
according to L'H\^{o}spital's rule. Varying the values of segment size $s$, we can determine the power-law relation between the function $F_q(s)$ and the size scale $s$,
\begin{equation}
  F_q(s) \sim {s}^{h(q)}.
  \label{Eq:hq}
\end{equation}
where $h(q)$ is the MF-DFA scaling exponent. When $q=2$, $h(2)$ is exactly the DFA scaling exponent $H^{\rm{DFA}}$.

\begin{figure}[htb]
  \centering
  \includegraphics[width=8cm]{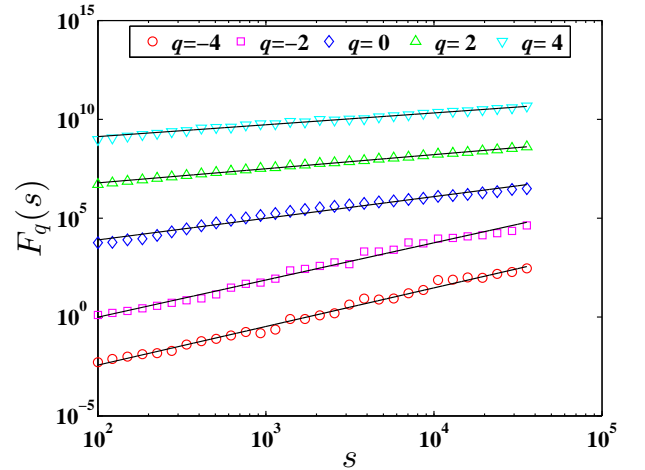}
  \caption{\label{Fig:MFDFA-FqS} Power-law dependence of the $q$-th detrended fluctuation function $F_q(s)$ of gaps on the buy LOBs against the window size $s$ of stock 000016. The solid lines are the best power-law fits to the data. The plots for $q=-2, 0, 2, 4$ have been translated vertically for better visibility.}
\end{figure}

Figure~\ref{Fig:MFDFA-FqS} illustrates the fluctuation functions $F_q(s)$ as a function of the scale sizes $s$ for the buy LOB of stock 000016. We observe that the functions $F_q(s)$ for different $q$ have nice power-law scaling with respect to $s$. Using least-squares linear regressions, we can obtain the scaling exponent $h(q)$. It is obvious that the scaling exponent $h(q)$ decreases with the order $q$. These observations are also present for price gap time series on the sell LOBs and for other stocks.

\begin{figure}[htb]
  \centering
  \includegraphics[width=8cm]{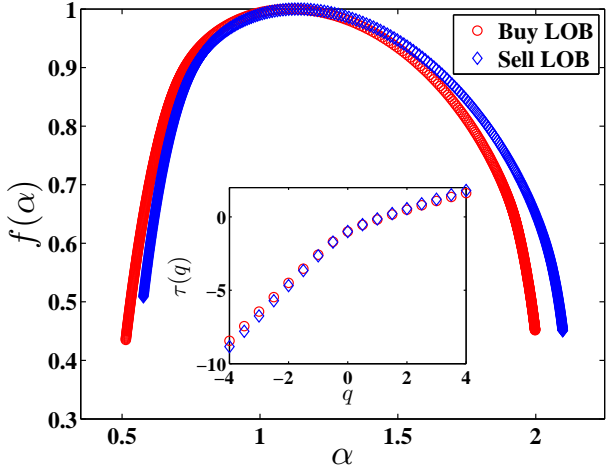}
  \caption{\label{Fig:MFDFA:f:alpha} Multifractal spectra $f(\alpha)$ of gap time series for both buy and sell LOBs of stock 000016. The inset presents the multifractal scaling exponents $\tau(q)$ as a function of $q$. }
\end{figure}

Based on the MF-DFA scaling exponent $h(q)$, we can calculate the multifractal scaling exponent $\tau(q)$ for one dimensional time series through
\begin{equation}
 \tau(q)=qh(q)-1,
 \label{Eq:MFDFA-tau}
\end{equation}
According to the standard multifractal formalism, if the scaling exponent $\tau(q)$ is a nonlinear function of $q$, the time series is considered to process multifractal nature. The scaling exponents $\tau(q)$ for both buy and sell LOBs of stock 000016 are presented in the inset of Fig.~\ref{Fig:MFDFA:f:alpha}. It is evident that the exponents $\tau(q)$ is nonlinear with regard to the order $q$, suggesting that the gap time series has multifractality.

Based on the Legendre transform, one obtains the singularity strength function $\alpha(q)$ and the multifractal spectrum $f(\alpha)$ \citep{Halsey-Jensen-Kadanoff-Procaccia-Shraiman-1986-PRA}:
\begin{equation}
\left\{
 \begin{array}{ll}
  \alpha(q)={\rm{d}}\tau(q)/{\rm{d}}q\\
  f(q)=q{\alpha}-{\tau}(q)
 \end{array}
\right..
 \label{Eq:MFDFA:f:alpha}
\end{equation}
Figure~\ref{Fig:MFDFA:f:alpha} shows the multifractal spectra $f(\alpha)$ as a function of the singularity strength $\alpha$ for both buy and sell LOBs of stock 000016.

The strength of multifractality can be quantitatively measured by the width of the multifractal spectrum, i.e., $\Delta\alpha=\alpha_{\rm{max}}-\alpha_{\rm{min}}$. A larger value of $\Delta\alpha$ corresponds to stronger multifractality. We obtain that $\Delta\alpha_{\rm{b}}=1.99-0.51=1.48$ for the buy LOB and $\Delta\alpha_{\rm{s}}=2.09-0.57=1.52$ for the sell LOB of stock 000016. Since the values of $\Delta\alpha$ are evidently larger than zero, it indicates that the gap series processes multifractality, which is consistent with the nonlinearity in the scaling exponent $\tau(q)$. The results for all the investigated stocks are listed in Table~\ref{Tb:MFDFA-dalpha}. The value of $\Delta\alpha_{\rm{b}}$ varies in the range $[1.25,~1.86]$ with an average value $\overline{\Delta\alpha}_{\rm{b}}=1.46\pm0.16$ for buy LOBs. The value of $\Delta\alpha_{\rm{s}}$ fluctuates in the range $[1.12,~1.91]$ with an average value $\overline{\Delta\alpha}_{\rm{s}}=1.46\pm0.20$ for sell LOBs. Since $\Delta\alpha$ are all larger than zero, the gap time series of the 26 stocks have multifractality.

\begin{table}[htp]
  \centering
  \caption{The width of multifractal spectra $\Delta\alpha$ of gaps for both buy and sell LOBs of 26 stocks based on the MF-DFA method. $\Delta\alpha_{\rm{SFL}}$ is the average spectrum width of 100 shuffled gap time series for each stock.}
  \medskip
  \label{Tb:MFDFA-dalpha}
  \centering
  \begin{tabular}{ccccccc|cccccccc}
  \hline\hline
  & \multicolumn{2}{c}{Buy LOB} && \multicolumn{2}{c}{Sell LOB} \\
  \cline{2-3} \cline{5-6}
  Stock & $\Delta\alpha_{\rm{b}}$ & $\Delta\alpha_{\rm{SFL}}$ && $\Delta\alpha_{\rm{s}}$ & $\Delta\alpha_{\rm{SFL}}$ \\
  \hline
    000001 & 1.27 & 0.24 && 1.18 & 0.23 \\
    000009 & 1.34 & 0.27 && 1.30 & 0.28 \\
    000012 & 1.48 & 0.27 && 1.43 & 0.26 \\
    000016 & 1.48 & 0.33 && 1.52 & 0.32 \\
    000021 & 1.37 & 0.24 && 1.43 & 0.23 \\
    000024 & 1.63 & 0.26 && 1.60 & 0.27 \\
    000066 & 1.44 & 0.29 && 1.42 & 0.28 \\
    000406 & 1.43 & 0.36 && 1.38 & 0.33 \\
    000429 & 1.41 & 0.38 && 1.57 & 0.40 \\
    000488 & 1.52 & 0.30 && 1.53 & 0.29 \\
    000539 & 1.86 & 0.26 && 1.89 & 0.27 \\
    000541 & 1.80 & 0.29 && 1.91 & 0.27 \\
    000550 & 1.40 & 0.25 && 1.41 & 0.24 \\
    000581 & 1.83 & 0.28 && 1.80 & 0.26 \\
    000625 & 1.39 & 0.21 && 1.34 & 0.20 \\
    000709 & 1.38 & 0.21 && 1.40 & 0.20 \\
    000778 & 1.57 & 0.30 && 1.57 & 0.29 \\
    000800 & 1.31 & 0.22 && 1.20 & 0.22 \\
    000825 & 1.25 & 0.31 && 1.12 & 0.34 \\
    000839 & 1.29 & 0.22 && 1.31 & 0.21 \\
    000858 & 1.50 & 0.28 && 1.49 & 0.27 \\
    000898 & 1.34 & 0.15 && 1.31 & 0.15 \\
    000917 & 1.49 & 0.24 && 1.62 & 0.23 \\
    000932 & 1.38 & 0.35 && 1.36 & 0.34 \\
    000956 & 1.33 & 0.28 && 1.32 & 0.24 \\
    000983 & 1.50 & 0.32 && 1.50 & 0.33 \\
  \hline\hline
 \end{tabular}
\end{table}

Multifractality may be influenced by the probability distribution of gaps. With the same test method as memory effect in Sec.~\ref{se:memory}, we first shuffle the gap series for 100 times, then calculate the width of the multifractal spectrum $\Delta\alpha_{\rm{SFL}}$ for each shuffled time series based on the MFDFA method. The average values from 100 shuffled series for each of the 26 stocks are also presented in Table~\ref{Tb:MFDFA-dalpha}. Since the values of $\Delta\alpha_{\rm{SFL}}$ are larger than zero, the distribution of gaps generates certain degree of spurious multifractality, which is probably due to the finite-size effect \citep{Zhou-2012-CSF}. For each gap time series, $\Delta\alpha_{\rm{b}}$ or $\Delta\alpha_{\rm{s}}$ is significantly greater than $\Delta\alpha_{\rm{SFL}}$, which means that the multifractal nature is intrinsic and not caused by the strong linear correlations quantified in Sec.~\ref{se:memory}.

\begin{figure}[htb]
  \centering
  \includegraphics[width=8cm]{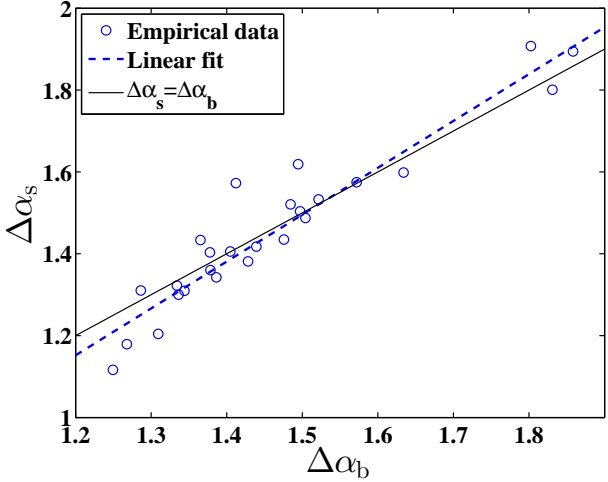}
  \caption{\label{Fig:MFDFA:dA:B:S} Relationship between the multifractal spectrum width $\Delta\alpha_{\rm{b}}$ for buy LOBs and $\Delta\alpha_{\rm{s}}$ for sell LOBs for the 26 stocks. The dash line is the linear fit to the empirical data and the solid line stands for the relation $\Delta\alpha_{\rm{s}}=\Delta\alpha_{\rm{b}}$.}
\end{figure}

We also compare the spectrum width of buy LOBs $\Delta\alpha_{\rm{b}}$ with the ones of sell LOBs $\Delta\alpha_{\rm{s}}$ for all the stocks which are illustrated in Fig.~\ref{Fig:MFDFA:dA:B:S}. There is an evident linear relationship and a robust linear regression gives
\begin{equation}
 \begin{split}
  \Delta\alpha_{{\rm{s}}}=&-0.2223~~+~~1.1452\Delta\alpha_{{\rm{b}}}~, \\
  &~~(0.0475)~~~~~~~(0.0000) \\
 \end{split}
 \label{Eq:RL-Deltaalpha}
\end{equation}
where the numbers in parentheses are the $p$-values of the coefficients. The coefficients obtained from the regression are significant at the 5\% level, especially for the coefficient of $\Delta\alpha_{{\rm{b}}}$ whose $p$-value is close to zero. These findings suggest that the multifractal nature changes from stock to stock and is thus not universal. Furthermore, the gap time series on the buy and sell LOBs share similar multifractal nature.

\section{Conclusion}
\label{se:conclusion}

In summary, we have investigated the statistical properties of the price gaps defined as the absolute logarithmic difference between the first occupied price level and second occupied price level on the LOBs based on order flow databases of 26 liquid A-share stocks. First, we study the probability distribution of gaps in both buy and sell LOBs, and find that the cumulative distribution function follows the power-law distribution in the tail with the average scaling exponent approaching 3.2. Then we investigate the memory effect of gap series using the detrending moving average (DMA) method and the detrended fluctuation analysis (DFA) method, respectively, and obtain similar results for all the stocks considered. The result indicates that the gap series of each stock processes strong long memory with the scaling exponent significantly larger than 0.5, and the probability distribution of gaps has no impact on its memory effect. Finally, we analyze the multifractal property of gap series applying the multifractal detrended fluctuation analysis (MF-DFA) method. It is evident that the gap series have multifractal nature for both buy and sell LOBs and the probability distribution of gaps has little effect on the multifractality.

Our work conducts a systemic investigation of the statistical properties of price gaps on the limit order books. These empirical findings not only deepen our understanding of the dynamics of liquidity, but also provide stylized facts for the calibration of agent-based computational models \citep{Li-Zhang-Zhang-Zhang-Xiong-2014-IS}. For instance, the seminal order-driven stock market model of \cite{Mike-Farmer-2008-JEDC} based on the statistical properties of order placement and order cancellation is able to reproduce the two main stylized facts, that is, the power-law tail distribution of returns and the absence of long-range correlations in return time series. However, it fails to reproduce the volatility clustering phenomenon. \cite{Gu-Zhou-2009-EPL} scrutinize the microscopic empirical rules of the model and improve the model by taking into account the long-term correlations in relative prices of submitted orders. They can thus reproduce the important stylized fact of volatility clustering. Along this line, one needs to check if the artificial stock market based on computational models has the unveiled stylized facts in the price gaps. If there are discrepancies between the artificial market and real markets, one can conclude that some ingredients are missing in the construction of the model and will need to modify or improve the model.

\section*{Acknowledgement}

This work was partly supported by National Natural Science Foundation of China (71101052, 71131007 and 71072007), Shanghai Rising Star (Follow-up) Program (11QH1400800), Program for Changjiang Scholars and Innovative Research Team in University (IRT1028), and the Fundamental Research Funds for the Central Universities.

%\bibliographystyle{elsarticle-harv}
%\bibliography{E:/Papers/Auxiliary/Bibliography}

\end{document}